\documentstyle[12pt]{article}

\font\twlgot =eufm10 scaled \magstep1 \font\egtgot =eufm8
\font\sevgot =eufm7 \font\twlmsb =msbm10 scaled \magstep1
\font\egtmsb =msbm8 \font\sevmsb =msbm7

\newfam\gotfam

\textfont\gotfam\twlgot \scriptfont\gotfam\egtgot
\scriptscriptfont\gotfam\sevgot

\newfam\msbfam
\textfont\msbfam\twlmsb \scriptfont\msbfam\egtmsb
\scriptscriptfont\msbfam\sevmsb
\def\Bbb{\protect\pBbb}
\def\pBbb{\relax\ifmmode\expandafter\Bb\else\typeout{You cann't use
Bbb in text mode}\fi}
\def\Bb #1{{\fam\msbfam\relax#1}}

\def\thebibliography#1{\section*{References}\list
  {[\arabic{enumi}]}{\settowidth\labelwidth{#1}\leftmargin\labelwidth
    \advance\leftmargin\labelsep
    \usecounter{enumi}}
    \def\newblock{\hskip .11em plus .33em minus .07em}
    \sloppy\clubpenalty4000\widowpenalty4000
    \sfcode`\.=1000\relax}

\def\op#1{\mathop{\fam0 #1}\limits}

\newcommand{\beq}{\begin{equation}}
\newcommand{\eeq}{\end{equation}}
\newcommand{\ben}{\begin{eqnarray}}
\newcommand{\een}{\end{eqnarray}}
\newcommand{\be}{\begin{eqnarray*}}
\newcommand{\ee}{\end{eqnarray*}}
\newcommand{\bea}{\begin{eqalph}}
\newcommand{\eea}{\end{eqalph}}

\newcommand{\rrq}{{\ol q}}

\newcommand{\dv}[1]{\dot{\rm{#1}}}

\newcommand{\vr}{\varrho}

\newcommand{\la}{\lambda}

\newcommand{\m}{\mu}

\newcommand{\g}{\gamma}
\newcommand{\G}{\Gamma}

\newcommand{\si}{\sigma}

\newcommand{\wt}{\widetilde}
\newcommand{\wh}{\widehat}
\newcommand{\ol}{\overline}
\newcommand{\dr}{\partial}
\newcommand{\ar}{\op\longrightarrow}
\newcommand{\ot}{\otimes}

\newcounter{eqalph}
\newcounter{equationa}
\newcounter{remark}
\newcounter{example}
\newcounter{theorem}
\newcounter{proposition}
\newcounter{lemma}
\newcounter{corollary}
\newcounter{definition}
\def\theremark{\arabic{remark}}
\def\thetheorem{\arabic{theorem}}

\def\thedefinition{\arabic{theorem}}

\newenvironment{prop}{\refstepcounter{theorem}
{\bf Proposition \thetheorem.}}{\medskip}

\newenvironment{defi}{\refstepcounter{theorem}
{\bf Definition \thedefinition.} }{}

\newenvironment{eqalph}{\stepcounter{equation}
\setcounter{equationa}{\value{equation}} \setcounter{equation}{0}

\begin{eqnarray}}{\end{eqnarray}
\setcounter{equation}{\value{equationa}}}

\newcommand{\mar}[1]{}

\begin{document}
\hbox{}

{\parindent=0pt

{\large \bf Relative non-relativistic mechanics}
\bigskip

{\sc G. Sardanashvily}

{\sl Department of Theoretical Physics, Moscow State University,
117234 Moscow, Russia}

\bigskip
\bigskip

\begin{small}

{\bf Abstract.} Dynamic equations of non-relativistic mechanics
are written in covariant-coordinate form in  terms of relative
velocities and accelerations with respect to an arbitrary
reference frame. The notions of the non-relativistic reference
frame, inertial force, free motion equation, and inertial frame
are discussed.

\end{small}

 }

\section{Introduction}

We consider second order dynamic equations in time-dependent
non-relativistic mechanics. A configuration space of
time-dependent non-relativistic mechanics is a smooth fibre bundle
\mar{gm360}\beq
\pi:Q\to\Bbb R. \label{gm360}
\eeq
A second order dynamic equation on this configuration space is a
closed subbundle of the second order jet bundle bundle $J^2Q\to
J^1Q$. Given bundle coordinates $(t,q^i)$ on $Q$ and the adapted
coordinates $(t,q^i,q^i_t,q^i_{tt})$ on $J^2Q$, such an equation
takes the coordinate form
\mar{z273}\beq
q^i_{tt}=\xi^i(t,q^j,q^j_t). \label{z273}
\eeq
We aim to bring this equation into the form (\ref{gm389})
maintained under bundle coordinate transformations and expressed
in relative velocities and accelerations with respect to an
arbitrary reference frame \cite{book98,book00}. Recall that a
reference frame in non-relativistic mechanics is defined as a
connection on the configuration bundle (\ref{gm360})
\cite{eche95,jmp00,massa,sard98}. The notions of an inertial
force, free motion equation, and inertial frame are discussed.

For instance, a dynamic equation  is said to be a free motion
equation if there exists a reference frame such that this equation
reads
\mar{z280}\beq
\ol q^i_{tt}=0. \label{z280}
\eeq
One can formulate the necessary criterion wether the dynamic
equation (\ref{z273}) is a free motion equation, but not the
sufficient one. With respect to an arbitrary reference frame, the
free motion equation (\ref{z280}) takes the form (\ref{m188}). One
can think of its right-hand side as being a general expression of
an inertial force in non-relativistic mechanics.

Note that Hamiltonian time-dependent mechanics with respect to an
arbitrary reference frame has been formulated
\cite{jmp07a,sard98}.

\section{Fibre bundles over $\Bbb R$}

Throughout the paper, a typical fibre $M$ of the fibre bundle $Q$
(\ref{gm360}) is an $m$-dimensional manifold. The base $\Bbb R$ of
$Q$ is parameterized by the Cartesian coordinates $t$ possessing
the transition functions $t'=t+$ const. It is provided with the
standard vector field $\dr_t$ and the standard one-form $dt$ which
are invariant under the coordinate transformations $t'=t+$ const.
The same symbol $dt$ also stands for any pull-back of the standard
one-form $dt$ onto fibre bundles over $\Bbb R$. Given bundle
coordinates $(t,q^i)$ on $Q$, we sometimes use the compact
notation $(q^\la)=(q^i,q^0=t)$ of them. Recall the notation
$d_t=\dr_t +q^i_t\dr_i +q^i_{tt}\dr^t_i$ of the total derivative,

Let us point out some peculiarities of fibre bundles and jet
manifolds over $\Bbb R$.

Since $\Bbb R$ is contractible, any fibre bundle over $\Bbb R$ is
obviously trivial. Its different trivializations
\mar{gm219}\beq
\psi: Q\cong  \Bbb R\times M \label{gm219}
\eeq
differ from each other in the projections $Q\to M$, while the
fibration $Q\to\Bbb R$ is once for all. Every trivialization
(\ref{gm219}) yields the corresponding trivialization of the jet
manifold $J^1Q\cong \Bbb R\times TM$. There is the  canonical
imbedding
\mar{z260}\beq
\la_1: J^1Q\hookrightarrow TQ, \qquad \la_1: (t,q^i,q^i_t) \mapsto
(t,q^i,\dot t=1, \dot q^i=q^i_t),\label{z260}
\eeq
of the affine jet bundle
\mar{f1}\beq
\pi^1_0:J^1Q\to Q \label{f1}
\eeq
to the tangent bundle $TQ$ of $Q$. Hereafter, we identify the jet
manifold $J^1Q$ with its affine image in $TQ$, modelled over the
vertical tangent bundle $VQ$ of the fibre bundle of $Q\to\Bbb R$.

A connection $\G$ on a fibre bundle $Q\to\Bbb R$ is defined as a
global section
\be
\G=dt\ot(\G = \dr_t + \G^i \dr_i)
\ee
of the affine jet bundle (\ref{f1}). In view of the morphism
$\la_1$ (\ref{z260}), it can be identified to a nowhere vanishing
horizontal vector field
\mar{a1.10}\beq
\G = \dr_t + \G^i \dr_i \label{a1.10}
\eeq
on $Q$ which is the horizontal lift of the standard vector field
$\dr_t$ on $\Bbb R$ by means of this connection. Conversely, any
vector field $\G$ on $Q$ such that $dt\rfloor\G =1$ defines a
connection on $Q\to\Bbb R$. The range of a connection $\G$
(\ref{a1.10}) is the kernel of the first order differential
operator
\mar{z279}\beq
 D_\G: J^1Q\op\to_Q VQ, \qquad \dot q^i\circ D_\G =q^i_t-\G^i, \label{z279}
\eeq
on $Q$ called the covariant differential of $\G$.

\begin{prop} \label{gena113} \mar{gena113}
Since a connection $\G$ on $Q\to\Bbb R$ is always flat, it defines
an atlas of local constant trivializations of $Q\to\Bbb R$ such
that the associated bundle coordinates $(t,\ol q^i)$ on $Q$
possess the time-independent transition functions, and $\G=\dr_t$
with respect to these coordinates. Conversely, every atlas of
local constant trivializations of the fibre bundle $Q\to\Bbb R$
determines a connection on  $Q\to\Bbb R$ which is equal to $\dr_t$
relative to this atlas \cite{book98,book00}.
\end{prop}

A connection $\G$ on a fibre bundle $Q\to \Bbb R$ is said to be
complete if the horizontal vector field (\ref{a1.10}) is complete.

\begin{prop} \label{complcon} \mar{complcon}
Every trivialization of a fibre bundle $Q\to \Bbb R$ yields a
complete connection on this fibre bundle. Conversely, every
complete connection $\G$ on $Q\to\Bbb R$ defines its
trivialization (\ref{gm219}) such that the vector field
(\ref{a1.10}) equals $\dr_t$ relative to the bundle coordinates
associated to this trivialization \cite{book98}.
\end{prop}

Let $J^1J^1Q$ be the repeated jet manifold of a fibre bundle
$Q\to\Bbb R$ (\ref{gm360}), provided with the adapted coordinates
$(t,q^i,q^i_t, q^i_{(t)},q^i_{tt})$. It possesses two affine
fibrations
\be
&& \pi_{11}:J^1J^1Q\to J^1Q, \qquad
q^i_t\circ\pi_{11}=q^i_t,\\
&& J^1_0\pi^1_0: J^1J^1Q\to J^1Q, \qquad q^i_t\circ
J^1_0\pi^1_0=q^i_{(t)},
\ee
which  are canonically isomorphic:
\mar{gm215}\beq
\pi_{11}\circ k= J^1_0\pi_{01},  \qquad
 q^i_t\circ k=q^i_{(t)}, \qquad q^i_{(t)}\circ k=q^i_t, \qquad
q^i_{tt}\circ k= q^i_{tt}. \label{gm215}
\eeq
The sesquiholonomic jet manifold $\wh J^2Q\subset J^1J^1Q$ and the
second order jet manifold $J^2Q\subset J^1J^1Q$ are isomorphic and
coordinated by $(t,q^i,q^i_t,q^i_{tt})$. The affine bundle
$J^2Q\to J^1Q$ is modelled over the vertical tangent bundle
\mar{gm217}\beq
V_QJ^1Q= J^1Q\op\times_QVQ\to J^1Q \label{gm217}
\eeq
of the affine jet bundle $J^1Q\to Q$. There is the imbedding
\mar{gm211}\ben
&& J^2Q \op\hookrightarrow^{\la_2} TJ^1Q
\op\hookrightarrow^{T\la_1} V_QTQ\cong T^2Q\subset TTQ,
\nonumber \\
&& \la_2: (t,q^i,q^i_t,q^i_{tt})\mapsto (t,q^i,q^i_t,\dot t=1,\dot
q^i=q^i_t,\dot q^i_t=q^i_{tt}), \label{gm211}
\een
where $(t,q^i,\dot t,\dot q^i,\dv t,\dv q^i, \ddot t, \ddot q^i)$
are the holonomic coordinates on the double tangent bundle  $TTQ$,
by $V_QTQ$ is meant the vertical tangent bundle of $TQ\to Q$, and
$T^2Q$ is a second order tangent space, given by the coordinate
relation $\dot t =\dv t$.

By a second order connection $\xi$ on a fibre bundle $Q\to\Bbb R$
(\ref{gm360}) is meant a connection on the jet bundle $J^1Q\to
\Bbb R$. Due to the imbedding (\ref{gm211}), it is represented by
a horizontal vector field
\be
\xi=\dr_t + \chi^i_t\dr_i + \xi^i \dr_i^t
\ee
on $J^1Q$ such that $\xi\rfloor dt=1$. A second order connection
which lives in $J^2Q\subset J^1J^1Q$ is called holonomic. It reads
\mar{a1.30}\beq
\xi=\dr_t + q^i_t\dr_i + \xi^i \dr_i^t. \label{a1.30}
\eeq
Its range is the kernel of the covariant differential
\mar{f3}\beq
D_\xi: J^1J^1Q\ar_{J^1Q} V_QJ^1Q, \qquad \dot q^i \circ D_\xi =0,
\qquad \dot q^i_t\circ D_\xi= q^i_{tt} - \xi^i. \label{f3}
\eeq

Every connection $\G$ on a fibre bundle $Q\to\Bbb R$ admits the
jet prolongation to a section $J^1\G$ of the affine bundle
$J^1\pi_0^1$ and, by virtue of the isomorphism $k$ (\ref{gm215}),
gives rise to the second order connection
\mar{gm217'}\beq
J\G\op= k\circ J^1\G: J^1Q\to J^1J^1Q,\qquad J\G=\dr_t +\G^i\dr_i
+ d_t\G^i\dr^t_i. \label{gm217'}
\eeq

\section{Dynamic equations}

A second order dynamic equation (or, simply, a dynamic equation)
on a fibre bundle $Q\to \Bbb R$, by definition, is the range of a
section of the jet bundle $J^2Q\to J^1Q$, i.e., a holonomic second
order connection $\xi$ (\ref{a1.30}) on $Q\to \Bbb R$
\cite{book98,book00}.  This equation is the kernel of the
covariant differential $D_\xi$ (\ref{f3}) given by the coordinate
equalities (\ref{z273}). The corresponding horizontal vector field
$\xi$ (\ref{a1.30}) is also called a dynamic equation. One can
easily find the transformation law
\mar{z317}\beq
q'^i_{tt} = \xi'^i, \qquad \xi'^i=(\xi^j\dr_j +
q^j_tq^k_t\dr_j\dr_k +2q^j_t\dr_j\dr_t +\dr_t^2)q'^i(t,q^j)
\label{z317}
\eeq
of a dynamic equation under coordinate transformations $q^i\to
q'^i(t,q^j)$. By a solution of the dynamic equation (\ref{z273})
is meant a section of $Q\to \Bbb R$ whose second order jet
prolongation lives in (\ref{z273}).

The fact that $\xi$ (\ref{a1.30}) is a curvature-free connection
places a limit on the geometric analysis of dynamic equations by
holonomic second order connections. Therefore, we consider the
relationship between the holonomic connections on the jet bundle
$J^1Q\to\Bbb R$ and the connections on the affine jet bundle
$J^1Q\to Q$ \cite{book98,book00}. The first order jet manifold of
$J^1Q\to Q$ is denoted by $J^1_QJ^1Q$.

Let $\g:J^1Q\to J^1_QJ^1Q$ be a connection on $J^1Q\to Q$. It
takes the coordinate form
\mar{a1.38}\beq
 \g=dq^\la\ot (\dr_\la + \g^i_\la \dr_i^t).
\label{a1.38}
\eeq
Let us consider the composite fibre bundle
\mar{gm361}\beq
J^1Q\to Q\to \Bbb R. \label{gm361}
\eeq
There is the canonical morphism
\be
\vr:  J^1_QJ^1Q \ni (q^\la,q^i_t,q^i_{\la t}) \mapsto
(q^\la,q^i_t,q^i_{(t)}=q^i_t,q^i_{tt}=q^i_{0t} +q^j_tq^i_{jt})\in
J^2Q.
\ee

\begin{prop}\label{gena51} \mar{gena51}
Any connection $\g$ (\ref{a1.38}) on the affine jet bundle
$J^1Q\to Q$ defines the second order holonomic connection
\mar{z281}\beq
\xi_\g=\vr\circ \g = \dr_t + q^i_t\dr_i +(\g^i_0
+q^j_t\g^i_j)\dr_i^t. \label{z281}
\eeq
\end{prop}

It follows that every connection $\g$ (\ref{a1.38}) on the affine
jet bundle $J^1Q\to Q$ yields the dynamic equation
\mar{z287}\beq
q^i_{tt}=\g^i_0 +q^j_t\g^i_j. \label{z287}
\eeq
on $Q\to\Bbb R$ which is the kernel, restricted to $J^2Q$, of the
vertical covariant differential
\mar{gm388}\beq
\wt D_\g: J^1J^1Q\to V_QJ^1Q, \qquad \dot q^i_t\circ\wt D_\g=
q^i_{tt} -\g^i_0 - q^j_t\g^i_j, \label{gm388}
\eeq
of a connection $\g$. Therefore, connections on the jet bundle
$J^1Q\to Q$ are called dynamic connections. A converse of
Proposition \ref{gena51} is the following.

\begin{prop}\label{gena52} \mar{gena52}
Any holonomic connection $\xi$ (\ref{a1.30}) on the jet bundle
$J^1Q\to \Bbb R$ yields the dynamic connection
\mar{z286}\beq
\g_\xi =dt\ot[\dr_t+(\xi^i-\frac12 q^j_t\dr_j^t\xi^i)\dr_i^t] +
dq^j\ot[\dr_j +\frac12\dr_j^t\xi^i \dr_i^t]. \label{z286}
\eeq
\end{prop}

It is readily observed that the dynamic connection $\g_\xi$
(\ref{z286}) possesses the property
\mar{a1.69}\beq
\g^k_i = \dr_i^t\g^k_0 +  q^j_t\dr_i^t\g^k_j \label{a1.69}
\eeq
which implies the relation $\dr_j^t\g^k_i = \dr_i^t\g^k_j$.
Therefore, a dynamic connection $\g$, obeying the condition
(\ref{a1.69}), is said to be {\sl symmetric}. \index{dynamic
connection symmetric} The torsion of a dynamic connection $\g$ is
defined as the tensor field
\mar{gm273}\beq
T=T^k_i \ol dq^i\ot\dr_k: J^1Q\to V^*Q\op\ot_Q VQ, \qquad T^k_i =
\g^k_i - \dr_i^t\g^k_0 - q^j_t\dr_i^t\g^k_j. \label{gm273}
\eeq
It follows at once that a dynamic connection is symmetric iff its
torsion vanishes. Let $\g$ be a dynamic connection (\ref{a1.38})
and $\xi_\g$ the corresponding dynamic equation (\ref{z281}). Then
the dynamic connection (\ref{z286}) associated to the dynamic
equation $\xi_\g$ takes the form
\be
\g_{\xi_\g}{}^k_i = \frac{1}{2} (\g^k_i + \dr_i^t\g^k_0 +
q^j_t\dr_i^t\g^k_j), \qquad \g_{\xi_\g}{}^k_0 = \xi^k -
q^i_t\g_{\xi_\g}{}^k_i.
\ee
It is readily observed that $\g = \g_{\xi_\g}$ iff the torsion $T$
(\ref{gm273}) of the dynamic connection $\g$ vanishes.

For instance, the affine jet bundle $J^1Q\to Q$ admits an affine
connection
\mar{z299}\beq
 \g=dq^\la\ot [\dr_\la + (\g^i_{\la 0}(q^\m)+ \g^i_{\la j}(q^\m)q^j_t)\dr_i^t].
\label{z299}
\eeq
This connection is symmetric iff $\g^i_{\la \m}=\g^i_{\m\la}$. One
can easily justify that an affine dynamic connection generates a
quadratic dynamic equation, and {\it vice versa}. Nevertheless, a
non-affine dynamic connection, whose symmetric part is affine,
also yields a quadratic dynamic equation.

\section{Reference frames}

From the physical viewpoint, a reference frame in non-relativistic
mechanics determines a tangent vector at each point of a
configuration space $Q$, which characterizes the velocity of an
"observer" at this point. This speculation leads to the following
notion of a reference frame in non-relativistic mechanics
\cite{eche95,jmp00,massa,sard98}.

\begin{defi}\label{gn10} \mar{gn10}
In non-relativistic mechanics, a reference frame is a connection
$\G$ on the configuration bundle $Q\to\Bbb R$.
\end{defi}

In accordance with this definition, the corresponding covariant
differential
\be
\dot q^i_\G\op= D_\G(q^i_t)= q^i_t-\G^i
\ee
determines the relative velocities with respect to the reference
frame $\G$.

By virtue of Proposition \ref{gena113}, any reference frame $\G$
on a configuration bundle $Q\to\Bbb R$ is associated to an atlas
of local constant trivializations, and {\it vice versa}. The
connection $\G$ reduces to $\G=\dr_t$ with respect to the
corresponding coordinates $(t,\rrq^i)$, whose transition functions
$\rrq^i\to \rrq'^i$ are independent of time. One can think of
these coordinates as  being also the reference frame,
corresponding to the connection $\G=\dr_t$. They are called the
adapted coordinates to the reference frame $\G$ or, simply, a
reference frame. In particular, with respect to the coordinates
$\rrq^i$ adapted to a reference frame $\G$, the velocities
relative to this reference frame are equal to the absolute ones
\be
D_\G(\rrq^i_t)={\dot \rrq}^i_\G=\rrq^i_t.
\ee
A reference frame is said to be complete if the associated
connection $\G$ is complete. By virtue of Proposition
\ref{complcon}, every complete reference frame defines a
trivialization  of a bundle $Q\to\Bbb R$, and {\it vice versa}.

Given a reference frame $\G$, one should solve the equations
\mar{gm300}\bea
&& \G^i(t, q^j(t,\rrq^a))=\frac{\dr q^i(t,\rrq^a)}{\dr t}, \label{gm300a}\\
&& \frac{\dr \rrq^a(t,q^j)}{\dr q^i}\G^i(t,q^j) +\frac{\dr
\rrq^a(t,q^j)}{\dr t}=0 \label{gm300b}
\eea
in order to find the coordinates $(t,\rrq^a)$ adapted to $\G$. Let
$(t,q^a_1)$ and $(t,q^i_2)$ be the adapted coordinates   for
reference frames $\G_1$ and $\G_2$, respectively. In accordance
with the equality (\ref{gm300b}),  the components $\G^i_1$ of the
connection $\G_1$ with respect to the coordinates $(t,q^i_2)$ and
the components $\G^a_2$ of the connection $\G_2$ with respect to
the coordinates $(t,q^a_1)$ fulfill the relation
\be
\frac{\dr q^a_1}{\dr q^i_2}\G^i_1 +\G^a_2=0.
\ee

Using the relations (\ref{gm300a}) -- (\ref{gm300b}), one can
rewrite the coordinate transformation law (\ref{z317}) of dynamic
equations as follows. Let
\mar{gm302}\beq
\rrq^a_{tt}=\ol\xi^a \label{gm302}
\eeq
be a dynamic equation on a configuration space $Q$, written with
respect to a reference frame $(t,\rrq^n)$. Then, relative to
arbitrary bundle coordinates $(t,q^i)$ on $Q\to\Bbb R$, the
dynamic equation (\ref{gm302}) takes the form
\mar{gm304}\beq
 q^i_{tt}=d_t\G^i +\dr_j\G^i(q^j_t-\G^j) -
\frac{\dr q^i}{\dr\rrq^a}\frac{\dr\rrq^a}{\dr q^j\dr
q^k}(q^j_t-\G^j) (q^k_t-\G^k) + \frac{\dr q^i}{\dr\rrq^a}\ol\xi^a,
\label{gm304}
\eeq
where $\G$ is the connection corresponding to the reference frame
$(t,\rrq^n)$. The dynamic equation (\ref{gm304}) can be expressed
in the relative velocities $\dot q^i_\G=q^i_t-\G^i$ with respect
to the initial reference frame $(t,\rrq^a)$. We have
\mar{gm307}\beq
 d_t\dot q^i_\G=\dr_j\G^i\dot q^j_\G -
\frac{\dr q^i}{\dr\rrq^a}\frac{\dr\rrq^a}{\dr q^j\dr q^k}\dot
q^j_\G \dot q^k_\G + \frac{\dr q^i}{\dr\rrq^a}\ol\xi^a(t,q^j,\dot
q^j_\G). \label{gm307}
\eeq
Accordingly, any dynamic equation (\ref{z273}) can be expressed in
the relative velocities $\dot q^i_\G=q^i_t-\G^i$ with respect to
an arbitrary reference frame $\G$ as follows:
\mar{gm308}\beq
d_t\dot q^i_\G= (\xi-J\G)^i_t=\xi^i-d_t\G, \label{gm308}
\eeq
where $J\G$ is the jet prolongation (\ref{gm217'}) of the
connection $\G$ onto $J^1Q\to\Bbb R$.

Let us consider the following particular reference frame $\G$ for
a dynamic equation $\xi$. The covariant differential of a
reference frame $\G$ with respect to the corresponding dynamic
connection $\g_\xi$ (\ref{z286}) reads
\mar{jp57}\ben
&&\nabla^\g\G = \nabla^\g_\la\G^k dq^\la\ot\dr_k :Q\to T^*Q\times
V_QJ^1Q,
\label{jp57}\\
&& \nabla^\g_\la\G^k = \dr_\la\G^k - \g^k_\la\circ\G. \nonumber
\een
A connection $\G$ is called a geodesic reference frame for the
dynamic equation $\xi$ if
\mar{gm310}\beq
\G\rfloor\nabla^\g \G= \G^\la(\dr_\la\G^k -
\g^k_\la\circ\G)=(d_t\G^i-\xi^i\circ \G)\dr_i=0. \label{gm310}
\eeq
It is readily observed that integral sections of a reference frame
$\G$ are solutions of a dynamic equation $\xi$ iff $\G$ is a
geodesic reference frame for $\xi$.

\section{Free motion equations}

We have called the dynamic equation (\ref{z273}) the free motion
equation if there exists a reference frame $(t,\ol q^i)$ on the
configuration bundle $Q$ such that this equation takes the form
(\ref{z280}). With respect to arbitrary bundle coordinates
$(t,q^i)$, a free motion equation reads
\mar{m188}\beq
 q^i_{tt}=d_t\G^i +\dr_j\G^i(q^j_t-\G^j) -
\frac{\dr q^i}{\dr\rrq^m}\frac{\dr\rrq^m}{\dr q^j\dr
q^k}(q^j_t-\G^j) (q^k_t-\G^k),  \label{m188}
\eeq
where $\G^i=\dr_t q^i(t,\ol q^j)$ is the connection associated to
the initial reference frame $(t,\ol q^i)$. One can think of the
right-hand side of the equation (\ref{m188}) as being a general
expression of an inertial force in non-relativistic mechanics. The
corresponding dynamic connection $\g_\xi$ on the affine jet bundle
$J^1Q\to Q$ is
\mar{gm366}\beq
\g^i_k=\dr_k\G^i  - \frac{\dr q^i}{\dr\rrq^m}\frac{\dr\rrq^m}{\dr
q^j\dr q^k}(q^j_t-\G^j), \qquad \g^i_0= \dr_t\G^i +\dr_j\G^iq^j_t
-\g^i_k\G^k. \label{gm366}
\eeq
Then, we come to the following criterion wether a dynamic equation
is a free motion equation \cite{book98}.

\begin{prop}
If $\xi$ is a free motion equation, then the curvature of the
corresponding dynamic connection $\g_\xi$ equals 0.
\end{prop}

This criterion fails to be sufficient. If the curvature of a
dynamic connection $\g_\xi$ vanishes, it may happen that
components of $\g_\xi$ equal 0 with respect to non-holonomic
bundle coordinates on the affine jet bundle $J^1Q\to Q$.

Note also that the dynamic connection (\ref{gm366}) is affine. It
follows that, if $\xi$ is a free motion equation, it is always
quadratic.

The free motion equation (\ref{m188}) is simplified if the
coordinate transition functions $\ol q^i\to q^i$ are affine in
coordinates $\ol q^i$. Then we have
\mar{m182}\beq
q^i_{tt}=\dr_t\G^i -\G^j\dr_j\G^i +2q^j_t\dr_j\G^i. \label{m182}
\eeq
The following shows that the free motion equation (\ref{m182}) is
affine in the coordinates $q^i$ and $q^i_t$ \cite{book98}.

\begin{prop}\label{gena115} \mar{gena115}
Let $(t,\rrq^a)$  be a reference frame on a configuration bundle
$Q\to \Bbb R$ and $\G$ the corresponding connection. Components
$\G^i$ of this connection with respect to another coordinate
system $(t,q^i)$ are affine functions of coordinates $q^i$ iff the
transition functions between the coordinates $\rrq^a$ and $q^i$
are affine.
\end{prop}

The geodesic reference frames for a free motion equation are
called inertial. They are $\G^i=v^i=$ const. By virtue of
Proposition \ref{gena115}, these reference frames define the
adapted coordinates
\mar{z320}\beq
\ol q^i =k^i_jq^j-v^it-a^i, \qquad k^i_j={\rm const.}, \qquad
v^i={\rm const.}, \qquad a^i={\rm const.}\label{z320}
\eeq
The  equation (\ref{z280}) keeps obviously its free motion form
under the transformations (\ref{z320}) between the geodesic
reference frames. It is readily observed that these
transformations are precisely the elements of the Galilei group.

\section{Relative acceleration}

It should be emphasized that, taken separately, the left- and
right-hand sides of the dynamic equation (\ref{gm308}) are not
well-behaved objects.  This equation can be brought into the
covariant form if we introduce the notion of a relative
acceleration.

To consider a relative acceleration with respect to a reference
frame $\G$,  one should prolong the connection $\G$ on the
configuration bundle $Q\to\Bbb R$ to a holonomic connection
$\xi_\G$ on the jet bundle $J^1Q\to\Bbb R$. Note that the jet
prolongation $J\G$ (\ref{gm217'}) of $\G$ onto $J^1Q\to\Bbb R$ is
not holonomic. We can construct the desired prolongation by means
of a dynamic connection $\g$ on the affine jet bundle $J^1Q\to Q$
\cite{book98}.

\begin{prop}\label{gn3} \mar{gn3}
Let us consider the composite bundle (\ref{gm361}). Given a frame
$\G$ on $Q\to\Bbb R$ and a dynamic connections $\g$ on $J^1Q\to
Q$, there exists a dynamic connection $\wt\g$ on $J^1Q\to Q$ with
the components
\mar{gm367}\beq
\wt \g^i_k=\g^i_k, \qquad \wt \g^i_0=d_t\G^i-\g^i_k\G^k.
\label{gm367}
\eeq
\end{prop}

We now construct a certain soldering form on the affine jet bundle
$J^1Q\to Q$, and add it to this connection. Let us apply
 the canonical projection $T^*Q\to V^*Q$ and then the imbedding
$\G:V^*Q\to T^*Q$ to the covariant differential (\ref{jp57}) of
the reference frame $\G$ with respect to the dynamic connection
$\g$.  We obtain the $V_QJ^1Q$-valued 1-form
\be
\si= [-\G^i(\dr_i\G^k - \g^k_i\circ\G)dt +(\dr_i\G^k -
\g^k_i\circ\G)dq^i]\ot\dr_k^t
\ee
on $Q$ whose pull-back onto $J^1Q$ is the desired soldering form.
The sum $\g_\G\op=\wt \g +\si$, called the frame connection, reads
\mar{jp68}\beq
\g_\G{}^i_0= d_t\G^i - \g^i_k\G^k -\G^k(\dr_k\G^i -
\g^i_k\circ\G),\qquad \g_\G{}^i_k= \g^i_k +\dr_k\G^i -
\g^i_k\circ\G. \label{jp68}
\eeq
This connection yields the desired holonomic connection
\be
\xi_\G^i= d_t\G^i +(\dr_k\G^i +\g^i_k - \g^i_k\circ\G)(q^k_t-\G^k)
\ee
on the jet bundle $J^1Q\to \Bbb R$.

 Let $\xi$ be a dynamic equation and
 $\g=\g_\xi$ the connection (\ref{z286}) associated to
$\xi$. Then one can think of the vertical vector field
\mar{gm369}\beq
a_\G\op=\xi-\xi_\G=(\xi^i-\xi_\G^i)\dr^t_i \label{gm369}
\eeq
 on the affine jet bundle $J^1Q\to Q$  as being a relative
acceleration with respect to the reference frame $\G$ in
comparison with the absolute acceleration $\xi$.

For instance, let us consider a reference frame $\G$ which is
geodesic for the dynamic equation $\xi$, i.e., the relation
(\ref{gm310}) holds. Then the relative acceleration with respect
to the reference frame $\G$ is
\be
(\xi-\xi_\G)\circ\G=0.
\ee

Let $\xi$ now be an arbitrary dynamic equation, written with
respect to coordinates $(t,q^i)$ adapted to the reference frame
$\G$, i.e., $\G^i=0$. In these coordinates, the relative
acceleration with respect to a reference frame $\G$  is
\mar{jp64}\beq
a^i_\G = \xi^i(t,q^j,q^j_t) -\frac12 q^k_t(\dr_k \xi^i - \dr_k
\xi^i\mid_{q^j_t=0}). \label{jp64}
\eeq
Given another bundle coordinates $(t, q'^i)$ on $Q\to\Bbb R$, this
dynamic equation takes the form (\ref{gm307}), while the relative
acceleration (\ref{jp64}) with respect to the reference frame $\G$
reads $a'^i_\G =\dr_jq'^i a^j_\G$. Then we can write a dynamic
equation (\ref{z273}) in the form which is covariant under
coordinate transformations, namely,
\mar{gm390}\beq
\wt D_{\g_\G} q^i_t = d_t q^i_t -\xi^i_\G=a_\G, \label{gm390}
\eeq
where $\wt D_{\g_\G}$ is the vertical covariant differential
(\ref{gm388}) with respect to the frame connection $\g_\G$
(\ref{jp68}) on the affine jet bundle $J^1Q\to Q$.

In particular, if $\xi$ is a free motion equation which takes the
form (\ref{z280}) with respect to a reference frame $\G$, then
\be
\wt D_{\g_\G} q^i_t=0
\ee
relative to arbitrary bundle coordinates on the configuration
bundle $Q\to\Bbb R$.

The left-hand side of the dynamic equation (\ref{gm390}) can also
be expressed in the relative velocities such that this dynamic
equation takes the form
\mar{gm389}\beq
d_t\dot q^i_\G -\g_\G{}^i_k\dot q^k_\G = a_\G,  \label{gm389}
\eeq
 which is the covariant form of the equation
(\ref{gm308}).

The concept of a relative acceleration is understood better when
we deal with a quadratic dynamic equation $\xi$, and the
corresponding dynamic connection $\g$ is affine. If a dynamic
connection $\g$ is affine, i.e.,
\be
\g^i_\la=\g^i_{\la 0} + \g^i_{\la k}q^k_t,
\ee
so is a frame connection $\g_\G$ for any frame $\G$:
\mar{gm386}\ben
&& \g_\G{}^i_{jk}=\g^i_{jk}, \nonumber\\
&& \g_\G{}^i_{0k} =\dr_k\G^i -\g^i_{jk}\G^j, \qquad \g_\G{}^i_{k0}
=\dr_k\G^i
-\g^i_{kj}\G^j, \label{gm368}\\
&& \g_\G{}^i_{00}= \dr_t\G^i-\G^j\dr_j\G^i +\g^i_{jk}\G^j\G^k.
\nonumber
\een
In particular,   we obtain
\be
\g_\G{}^i_{jk}=\g^i_{jk}, \qquad
\g_\G{}^i_{0k}=\g_\G{}^i_{k0}=\g_\G{}^i_{00}=0
\ee
relative to the coordinates adapted to a reference frame $\G$. A
glance at the expression (\ref{gm368}) shows that, if a dynamic
connection $\g$ is symmetric, so is a frame connection $\g_\G$.
Thus, we come to the following.

\begin{prop} \label{gn5} \mar{gn5}
If a dynamic equation $\xi$ is quadratic, the relative
acceleration $a_\G$ (\ref{gm369}) is always affine, and it admits
the decomposition
\mar{gm371}\beq
a_\G^i= -(\G^\la\nabla^\g_\la\G^i + 2\dot q^\la_\G
\nabla^\g_\la\G^i), \label{gm371}
\eeq
where $\g=\g_\xi$ is the dynamic connection (\ref{z286}), and
\be
\dot q^\la_\G=q^\la_t -\G^\la, \qquad q^0_t=1, \qquad \G^0=1,
\ee
is the relative velocity with respect to the reference frame $\G$.
\end{prop}

Note that the splitting (\ref{gm371}) provides a  generalized
Coriolis theorem.  In particular, the well-known analogy between
inertial and electromagnetic forces is restated. Proposition
\ref{gn5} shows that this analogy can be extended to an arbitrary
quadratic dynamic equation.

\end{document}